\documentclass[showpacs,twocolumn,superscriptaddress]{revtex4}

\usepackage{doi}%<----------
\usepackage{hyperref}
\hypersetup{
  colorlinks=true,        % false: boxed links; true: colored links
  linkcolor=blue,         % color of internal links
  citecolor=cyan,         % color of links to bibliography
}

\usepackage{graphicx}% Include figure files
\usepackage{dcolumn}% Align table columns on decimal point
\usepackage{bm}% bold math
\usepackage{color}

\usepackage{amsmath}
\usepackage{amssymb}

\begin{document}

\title{Geodesic circular orbits sharing the same orbital frequencies in the black string spacetime 
 }

\author{Sanjar Shaymatov}
\email{sanjar@astrin.uz}

\affiliation{Institute for Theoretical Physics and Cosmology, Zheijiang University of Technology, Hangzhou 310023, China}\affiliation{Ulugh Beg Astronomical Institute, Astronomicheskaya
33, Tashkent 100052, Uzbekistan}
%\affiliation{National University of Uzbekistan, Tashkent 100174, Uzbekistan}
\affiliation{Tashkent Institute of Irrigation and Agricultural Mechanization Engineers,\\ Kori Niyoziy 39, Tashkent 100000, Uzbekistan }

\author{Farruh Atamurotov}
\email{atamurotov@yahoo.com}

\affiliation{Inha University in Tashkent, Tashkent, 100170, Uzbekistan}
\affiliation{Ulugh Beg Astronomical Institute, Astronomicheskaya
33, Tashkent 100052, Uzbekistan}

\date{\today}
\begin{abstract}
{ Here we consider isofrequency pairing of geodesic orbits that share the same three orbital frequencies associated with $\Omega^{\hat{r}}$, $\Omega^{\hat{\varphi}}$, and $\Omega^{\hat{\omega}}$ in a particular region of parameter space around black string spacetime geometry. We study the effect arising from the extra compact spatial dimension on the isofrequency pairing of geodesic orbits and show that such orbits would occur and become in a cylindrical manner in the allowed region as particles move along the black string. We find that the effect due to the extra dimension leads to an increase in the number of the isofrequency pairing of geodesic orbits. We also find that isofrequency pairing of geodesic orbits in the region of parameter space can not be realized beyond the critical value $J_{cr} \approx 0.096$ of the conserved quantity of the motion due to the extra compact spatial dimension for given parameters of the black string.}
 
\end{abstract}
\maketitle

%\section{Introduction}

\section{Introduction}
\label{introduction}

{In general relativity, the black holes are regarded as one of the most interesting and fascinating objects due to their geometric and remarkable gravitational properties. However, the point to be noted is that their existence had been predicted as a potential tool in explaining observational phenomena until the discovery of the gravitational waves detected by LIGO and Virgo scientific collaborations \cite{Abbott16a,Abbott16b} and  as well as the first direct shadow of supermassive black hole at the center of galaxy 87 by the Event Horizon Telescope (EHT) collaboration~\cite{Akiyama19L1,Akiyama19L6}. These observations opened a qualitatively new stage to reveal black hole's unknown properties and test remarkable nature of the  background geometry around black hole's horizon irrespective of the fact that there are still fundamental problems that general relativity faces, i.e., the occurrence of singularity, spacetime quantization, etc. In this framework, the motion of test particles in the strong gravitational field regime has been a productive field of study for several years~\cite{Herrera00,Herrera05,Kovar08,Kovar10,
Shaymatov18a,Dadhich18,Pavlovic19,Shaymatov19b,Bini12,
Toshmatov19d,Haydarov20,Stuchlik20,Shaymatov20egb,Shaymatov20b}. On the other hand, there is an extensive body of work devoted to understand the nature of radiative inspirals around black holes as a source of gravitational waves and binary systems ~\cite{Carter68,Bardeen72,Drasco04,Barack11,Warburton13,Shaymatov14,Kunst15,Osburn16,Lewis17}.    } 

{ It is worth noting that experiments and observations always allow to test the nature of the geometry through alternative theories of modified gravity in the strong field regime. The spactime geometry that can depart from the spherically symmetric Schwarzschild black hole was considered by Grunau and Khamesra~\cite{Grunau13}. This spacetime metric describes a five dimensional black string due to the extra compact dimension added to the well known Schwarzschild metric, and thus it is one of reasons why this solution is interesting object to investigate. The motion of test particles in the different gravity string models has been investigated thoroughly by several authors~\cite{Galtsov89,Chakraborty96,Ozdemir03,Ozdemir04,Hackmann10a,Hackmann10b}. Also, the geodesic motion of test particles in the black string spacetime for both rotating and non-rotating cases has been studied in detail~\cite{Grunau13}.  The geodesic motion of colliding particles was also studied in the vicinity of black string spacetime~\cite{Tursunov13}. } 

{The theoretical investigation of isofrequency pairing of geodesic circular orbits has been considered in vicinity of Schwarzschild and Kerr black holes, i.e., the region of the parameter space where such geodesic orbits occur has been discussed by providing an intuitive explanation of their occurrence.  The above investigation was addressed by taking an external asymptotically uniform magnetic field in the Schwarzschild spacetime, and then it turned out that the surface of allowed region where isofrequency pairing orbits occur decreases due to the magnetic field~~\cite{Shaymatov14}.  It was later extended to the spinning particles in Schwarzschild-de Sitter spacetime~\cite{Kunst15}. In the present paper we consider isofrequency pairing of geodesic orbits around black string spacetime geometry, as shown by the line element in \cite{Grunau13}, and we study the effect of the extra compact spatial dimension $\omega$ on the isofrequency pairing of geodesic spiral orbits sharing the same three orbital frequencies. }
 
{In this paper we consider a new frequency
$\Omega^{\hat{\omega}}$ arisen from the extra compact spatial dimension, which is associated with the motion in the extra direction along the black string. Generally, the bound orbits are confined to the interior of a compact special spiral given by $h_{z1}(e=0)\leq h_z \leq h_{z2} (e=1)$; frequencies, in particular, $\Omega^{\hat{r}}$ and $\Omega^{\hat{\theta}}$ are considered as "libration"-type frequencies associated with the radial and longitudinal periods, while $\Omega^{\hat{\varphi}}$ is a "rotation"-type frequency. In the case of black string spacetime, new frequency $\Omega^{\hat{\omega}}$ can be defined by a "libration-rotation" type frequency which involves two kinds of type frequencies, as described above. For the sake of clarity, we will further focus on the motion of test particles at the equatorial plane (i.e. $\theta=\pi/2$), and hence we omit $\Omega^{\hat{\theta}}$ as it loses its meaning. Then bound geodesic orbits around black string spacetime are respectively characterized by three orbital frequencies, i.e. $\Omega^{\hat{r}}$,  $\Omega^{\hat{\varphi}}$, and $\Omega^{\hat{\omega}}$ associated with the periodic motions. Thus, these orbits are considered as triperiodic orbits around the black string spacetime. 
In doing so we predict that an infinite number of pairs of such orbits may occur in a region of parameter space around the black string spacetime where physically distinct orbits possessing $\mathcal{E}$, ${\cal L}$, and ${\cal J}$ share the same orbital frequencies associated with $\Omega^{\hat{\varphi}}$, $\Omega^{\hat{r}}$, and
$\Omega^{\hat{\omega}} $, thereby referring to the “isofrequency” bound geodesic orbits synchronized in phase $\varphi$  while passing periastra at
the same time. To consider isofrequency pairing of geodesic orbits in the vicinity of black string spacetime may in principle play an important role in an astrophysical scenario in understanding the nature of radiative inspirals around black holes. }

{The paper is organised as follows: In Sec.\
\ref{Sec:metric} we briefly describe the metric for black string and provide the geodesic equation of motion around black string spacetime. We further study the effect of the extra compact spatial dimension on the isofrequency pairing of geodesic spiral orbits in Sec.~\ref{Sec:isofrequency}. We end up with conclusion in Sec.~\ref{Sec:Conclusion}. Throughout the paper we use a system of geometric units in which $G=c=1$. }

%%%%%%%%%%%%%%%%%%%%%%%%%%%%%
\section{The metric and the geodesic equations of motion }
\label{Sec:metric}

The metric describing five dimensional black string spacetime with an extra compact spatial dimension $\omega$ added to the Schwarzschild metric in the Boyer-Lindquist coordinates $(t, r,
\theta, \varphi, \omega)$ is given by~\cite{Grunau13}
\begin{eqnarray}\label{metric}
ds^{2} & = & -\bigg(1-\frac{2M}{r}\bigg) dt^{2}
+ \bigg(1-\frac{2M}{r}\bigg)^{-1} dr^{2} \nonumber\\
&&+~ r^{2}(d\theta^{2}+ \sin^{2}d\varphi^{2})+d\omega^{2}  \,\, ,\nonumber\\
\end{eqnarray}
with $M$ being the total mass of black string and $\omega$ corresponding yo the extra compact spatial dimension.  

The radial motion of geodesic test particles in the
equatorial plane of black string spacetime satisfies the following equation
\begin{eqnarray}\label{Eff}
\dot{r}^2={\cal E}^2-V_{eff}(r;\cal L,\mathcal{J}),
\end{eqnarray}
where 
\begin{eqnarray}\label{Eff1}
V_{eff}(r;\mathcal
L,\mathcal J)=\left(1-\frac{2M}{r}\right)\left(1+\mathcal{J}^{2}+\frac{{\mathcal
L}^2}{r^2}\right)\, ,
\end{eqnarray}
with $\mathcal{E}={E}/{m}$, ${\cal L} = {L}/{m}$, and ${\mathcal J} = {J}/{m}$ being the conserved constants per unit mass, related to the particle's specific energy, angular momentum, and conserved quantity due to $\omega$, respectively. It is well known that bound orbits exist in the case when $\mathcal{L}>2\sqrt{3}M$
with $2\sqrt{2}/3<\mathcal{E}<1$ for the Schwarzschild black hole. Whereas, in the case of black string, bound orbits only exist when $\mathcal{L}>2\sqrt{3}(1+\mathcal{J}^2)^{1/2}M$ with $2\sqrt{2}(1+\mathcal{J}^2)^{1/2}/3<\mathcal{E}<(1+\mathcal{J}^2)^{1/2}$ is satisfied, and the motion of particles is then restricted by the turning points being labeled as the periastron $r_{\rm p}$ and the apastron $r_{\rm a}$, respectively.

The conditions $V_{eff}(r_{p},\mathcal L,\mathcal{J})=V_{eff}(r_{a},\mathcal L,\mathcal{J})={\cal E}^{2}$
lead to the specific energy $\mathcal{E}$ and angular momentum $\mathcal{L}$ of the particle, given in terms of $p$ and $e$. Note that parameters $p$ and $e$ respectively measure the size of the orbit and 
its degree of noncircularity~\cite{Darwin61}.
Then we have 
\begin{eqnarray}
    \mathcal{E}^2 & = &\frac{(p-2-2e)(p-2+2e)}{p(p-3-e^2)}(1+\mathcal{J}^2),\nonumber\\
     &&\mathcal{L}^2  =  \frac{p^2M^2}{p-3-e^2}(1+\mathcal{J}^2).
    \label{eq:Energy_Ang_Mom}
\end{eqnarray}

Following Cutler \emph{et al.} \cite{Cutler94}, we turn to
the integration of the geodesic equations in terms of $t(r)$, $\varphi(r)$, and $\omega(r)$ coordinates 
\begin{eqnarray}\label{tr}
t(r) = \mathcal{E} \int \frac{r dr}{
(r-2M)\left[\mathcal{E}^2-V_{eff}(r;\cal L,\mathcal{J})\right]^{1/2}},
\end{eqnarray}
\begin{eqnarray}\label{fr}
\varphi(r) = \mathcal{L} \int
\frac{dr}{r^2\left[\mathcal{E}^2-V_{eff}(r;\cal L,\mathcal{J})\right]^{1/2}},
\end{eqnarray}
and
\begin{eqnarray}\label{omega}
\omega(r) &=& \mathcal{J} \int \frac{dr}{\left[\mathcal{E}^2
-V_{eff}(r;\cal L,\mathcal{J})\right]^{1/2}}.
\end{eqnarray}

Using Eqs.~ (\ref{eq:Energy_Ang_Mom})--(\ref{omega}) let us then turn to the following expressions with the parameter $\chi$ related to the coordinates $t$, $\varphi$, and $\omega$ as follows
\begin{eqnarray}
    \frac{dt}{d\chi} &=& \frac{Mp^2[(p-2)^2-4e^2]^{1/2}(p-6-2e\cos\chi)^{-1/2}}{(p-2-2e\cos\chi)(1+e\cos\chi)^2}\, ,
    \label{eq:dt_dchi} \nonumber\\
\end{eqnarray}
\begin{eqnarray}
    \frac{d\varphi}{d\chi} &=& \frac {p^{1/2}}{(p-6-2e\cos\chi)^{1/2}}\, , \label{eq:dphi_dchi}
\end{eqnarray}
\begin{eqnarray}
    \frac{d\omega}{d\chi} &=& \frac {\mathcal{J}M p^{3/2}(p-3-e^2)^{1/2} }{(1+\mathcal{J}^{2})^{1/2}(p-6-2e\cos\chi)^{1/2}(1+e\cos\chi)^2}\,
    .
    \label{eq:omega_dchi} \nonumber\\
\end{eqnarray}

The radial period and frequency are defined by \cite{Cutler94,Warburton13,Shaymatov14} 
\begin{eqnarray}
T^{\hat{r}}         = \int^{2\pi}_0 \frac{dt}{d\chi}\,d\chi\,,
~~~~~\Omega^{\hat{r}} = \frac{2\pi}{T^{\hat{r}}} \, .
\label{eq:T_r}
\end{eqnarray}
%\label{eq:r_frequency}
%
The azimuthal frequency of the orbit will then have the form as
\begin{eqnarray}
\Omega^{\hat{\varphi}} =
\frac{1}{T^{\hat{r}}}\int_0^{T^{\hat{r}}}\frac{d\varphi}{dt}dt=\frac{\Delta\varphi}{T^{\hat{r}}}\,
,  \label{eq:phi_frequency}
\end{eqnarray}
with azimuthal phase $\Delta\varphi$, which is given by     
\begin{eqnarray}
\Delta\varphi   &=&  \int^{2\pi}_0 \frac{d\varphi}{d\chi}\,d\chi = \nonumber\\
&=&
  \int_{0}^{2\pi}\frac {\sqrt{p}}{\sqrt{p-6-2e\cos\chi}}\, d\chi = 4 \sqrt{ \frac{p}{\epsilon}}\,
K\left(-\frac{4e}{\epsilon}\right)\, , \nonumber\\
 \label{eq:DeltaPhi}
\end{eqnarray}
where $\epsilon = p-p_s(e)=p-6-2e$ and
$K(x)=\int_{0}^{\pi/2}d\theta(1-x\sin^2\theta)^{-1/2}$ is the
complete elliptic integral of the first kind.

The new extra compact frequency arising from  the effect of the compact
spatial dimension $\omega$ of the orbit can be defined by
\begin{eqnarray}
\Omega^{\hat{\omega}} =
\frac{1}{T^{\hat{r}}}\int_0^{T^{\hat{r}}}\frac{d\omega}{dt}dt=\frac{\Delta\omega}{T^{\hat{r}}}\,
,  \label{eq:omega_frequency}
\end{eqnarray}
where $\Delta\omega$ is the new extra phase accumulated over time
interval $T^{\hat{r}}$.
\section{Isofrequency pairing orbits in black string spectime geometry}\label{Sec:isofrequency}

Now we plan to investigate the isofrequency and bound orbits of
the particles in the vicinity of the gravitational object in the
black string spacetime. We consider orbits lying extremely close
to the separatrix $\epsilon$ in order to estimate the divergent
quantities of the azimuthal and the new extra phases, the
radial period, and their ratio in Eqs.\ (\ref{eq:phi_frequency})
and (\ref{eq:omega_frequency}). Considering the near-separatrix
analytic expansions we obtain the corresponding expression for
$\Delta\omega$ as
\begin{eqnarray}
    \Delta\omega   &=&   \frac{\mathcal{J}M(3+2e-e^2)^{1/2}}{(1+\mathcal{J}^2)^{1/2}(1+2e)}\sqrt{\frac{(6+2e)^3}{e}}\ \nonumber\\
                &&
 \times\left[1+  \mathcal{O}\left(\frac{\epsilon}{4 e}\right)\right]\ln\left(\frac{64e}{\epsilon}\right)\, .
 \label{deltaomega}
\end{eqnarray}

The previous investigations suggest that the
occurrence of isofrequency pairing of geodesic orbits is strongly related to the presence of boundary regions referred as {\it separatrix} and {\it circular-orbit duals} (COD), and each and every circular orbit in the open range between the separatrix and singular curve and between the singular curve and COD as well has a dual isofrequency pairing. Note that the existence of the {\it separatrix} plays an important role for occurrence of these dual orbits in this isofrequency scenario. To consider the singular curve and the circular orbit duals in the particular region of the gravitational object is particularly important, and the COD can also play an important role as a boundary region allowing to keep pairs of isofrequency orbits exist in this region~\cite{Warburton13,Shaymatov14,Kunst15}.

Let us then come to investigate the effect of the extra compact spatial dimension on the occurrence of the isofrequency pairing of geodesic orbits in black string spacetime. Based on the above discussions and Eqs.~(\ref{eq:T_r})-(\ref{eq:DeltaPhi}) we keep the result for which
the range of frequencies of any pairs of orbits on $\Omega^{\hat{\varphi}}=(M/r_b^3)^{1/2}$ in the black
string is given by
\begin{eqnarray}
\tilde\Omega^{\hat{\varphi}}_{e=0}=0.062~<
~\tilde\Omega^{\hat{\varphi}}~<
~\tilde\Omega^{\hat{\varphi}}_{e=1}=0.125\, .
\end{eqnarray}
Also we determine the values of radial and azimuthal frequencies numerically and tabulate their values in Table~\ref{table1}.
\begin{table}
\caption{Numerical values of $\Omega^{\hat{r}}$
and $\Omega^{\hat{\varphi}}$ of isofrequency pairing of geodesic
spiral orbits around the black string spacetime for the different values of eccentricity $e$.} \label{table1}
\begin{tabular}{l l l l l l l l }
\hline\hline\noalign{\smallskip}
$e$& $0.1$ & $0.3$ & $0.4$ & $0.5$ & $0.6$   \\
$\Omega^{\hat{\varphi}}$ & 0.063 &0.068  &  0.073 & 0.079 & 0.091  \\
$\Omega^{\hat{r}}$ & 0.0069 &0.0124  &  0.0152 & 0.0181 & 0.0226
\\ \hline\noalign{\smallskip}
$e$& $0.7$ & $0.8$ & $0.89$ & $0.96$ & 1   \\
$\Omega^{\hat{\varphi}}$ & 0.103 &0.109  &  0.115 & 0.121 & 0.125
\\
$\Omega^{\hat{r}}$ & 0.0272 &0.0304  &  0.0334 & 0.0362 & 0.0380
\\ \hline\hline\noalign{\smallskip}
\end{tabular}
\end{table}
In Table~\ref{table2}, we show numerical values of the frequency
$\tilde\Omega^{\hat{\omega}}$ of the bound orbits around the
black string.  As can be seen from Table~\ref{table2} the value of the frequency $\tilde\Omega^{\hat{\omega}}$ arisen from the extra compact spatial dimension is increasing with an increase in value of the forth conserved quantity $J$, but the height of the horizontal direction along black string is decreasing for given values of eccentricity.
However, numerical analysis leads to the fact that when the value of the fourth conserved quantity attains $\mathcal J_{cr}\approx 0.096$ geodesic spiral orbits are not allowed all through to occur along the black string. Thus, the pairs of geodesic circular spiral orbits of particles with the same radial $\tilde\Omega^{\hat{r}}$, azimuthal $\tilde\Omega^{\hat{\varphi}}$, and new  extra
$\tilde\Omega^{\hat{\omega}}$ frequencies occur in the particular region of the gravitational object under this critical value irrespective of the fact that these orbits are physically distinct with the different values of conserved quantities (i.e. $\mathcal{E}$, ${\mathcal L}$, and ${\mathcal J}$ ). Then the pairs of these spiral orbits of particles with three frequencies associated with ($\tilde\Omega^{\hat{r}}$, $\tilde\Omega^{\hat{\varphi}}$, $\tilde\Omega^{\hat{\omega}}$) would move and oscillates in a cylindrical manner along the black string. 

It turns out that the decrease in the value of conserved fourth constant $\mathcal J$ makes the above mentioned cylindrical region become larger around the black string. As a result, this causes, in turn, to play an important role not in restricting the motion of the particles along the black string and in increasing the amount of isofrequency pairs of
geodesic spiral orbits in the region between the {\it separatrix}
and {\it circular-orbit duals} as well as between horizontal length $h_{z}$, i.e. in the particular area of the black string. Hence, one can keep in mind that there would exist an infinite number of pairs of circular spiral orbits having the same three orbital frequencies, in spite of physically distinct obits with three conserved quantities, in
the particular region around the black string. This result, in the presence of extra compact spatial dimension, leads to the conclusion that the occurrence of these spiral orbits would be more particularly important in understanding and explaining the behavior of black string spacetime as well as the nature of the radiative inspirals as a source of gravitational waves or binary systems. 
\begin{table}
\caption{Numerical values for the frequency
$\Omega^{\hat{\omega}}$ of isofrequency pairing of geodesic spiral orbits around the black string for the different values of the fourth
conserved parameter $\mathcal J$.} \label{table2}
\begin{tabular}{l l l l l l l l}
\hline\hline\noalign{\smallskip}
$$ & $\mathcal J$  & $0.01$ &   $0.03$   & $0.05$  & $0.07$   & $0.09$ & $0.096$   \\[0.5ex]
\hline
$e=0$ &$\tilde\Omega^{\hat{\omega}}$ & 0.007 & 0.021 &  0.035 &
 0.049 &0.063  & \, \, \,-\\
$$ &$r $ &27.17 & 13.05 & 9.29 & 7.43 &  6.29 &
\\
$$ &$h_{z} $ &26.51 & 11.59  & 7.09  & 4.38 &  1.89 &
\\
\hline
$e=1$ & $\tilde\Omega^{\hat{\omega}} $ &0.006 & 0.019 & 0.033 &
0.046  & 0.059 & \, \, \,-
\\
$$ &$r $ & 28.25& 13.57 & 9.66 & 7.73 & 6.55 & 
\\
$$ &$h_{z} $ &27.53 & 11.98 &  7.25 &  4.36 &  1.45 &
\\
\noalign{\smallskip}\hline\hline
\end{tabular}

\end{table}
\section{Conclusions}
\label{Sec:Conclusion}

We have considered isofrequency pairing of geodesic spiral orbits in the black string spacetime, which recovers the Schwarzschild one in the case of vanishing extra compact spatial dimension $\omega$. Thus, in an astrophysical scenario, it is worth investigating these geodesic orbits in the vicinity of black string as it allows to test the effects arising from the five dimensions.  
Our analysis suggests that the extra compact dimension leads to an increase in the number of pairs of circular spiral orbits.  These pairs of orbits having the same three orbital frequencies (i.e. $\Omega^{\hat{r}}$, $\Omega^{\hat{\varphi}}$, and
$\Omega^{\hat{\omega}}$) occur in the particular region and oscillate in a cylindrical manner around the black string in spite of the fact that they are completely physically distinct orbits possessing different conserved quantities (i.e. $\mathcal{E}$, ${\mathcal L}$, and ${\mathcal J}$). Also we showed that pairs of geodesic spiral orbits are not allowed all through to occur in the black string vicinity in the case when the fourth conserved quantity attains $\mathcal J \approx 0.096$. 

The obtained results suggest that an infinite number of pairs of such orbits may occur in the particular region around the black string and that it would be particularly important in explaining not only the possibility of occurrence of pairs of geodesic spiral orbits around the black holes, but also the nature of the black string. These theoretical results and discussions would be useful to the possible interpretation of astrophysical observations, and they can help to provide an information on the validity of alternative models to black holes and may also explain the nature of the radiative inspirals as isofrequency paring of geodesic spiral orbits can give the same information as such inspiral objects.

\section*{Acknowledgments}
The authors acknowledge the support of the Uzbekistan Ministry for Innovative Development Projects No.
VA-FA-F-2-008 and No. MRB-AN-2019-29.

\appendix

\bibliographystyle{apsrev4-1}  
\bibliography{gravreferences}

\end{document}